\documentclass[twocolumn,preprintnumbers,amsmath,amssymbn,reprint,superscriptaddress,nofootinbib]{revtex4-2}

\usepackage{bbold}
\usepackage{mathrsfs}
\usepackage[utf8]{inputenc} 
\usepackage[hidelinks]{hyperref}
\usepackage{graphicx}
\usepackage[dvipsnames]{xcolor}

\DeclareMathOperator{\Tr}{Tr}

\newcommand{\ket}[1]{| #1\rangle}
\newcommand{\bra}[1]{\langle #1|}
\newcommand{\braket}[2]{\langle #1|#2\rangle}
\newcommand{\ident}{\mathbb{1}}
\newcommand{\rp}{\ensuremath{P}}
\newcommand{\rhoi}{\ensuremath{\hat{\rho}}_I}

\newcommand{\vis}[1]{\ensuremath{\mathcal{V}_{#1}}}
\newcommand{\stokes}[1]{\ensuremath{\mathcal{S}_{#1}}}
\newcommand{\stokespar}[1]{\ensuremath{\textup{S}_{#1}}}

\newcommand{\visx}{\ensuremath{\stokes{x}}}
\newcommand{\visy}{\ensuremath{\stokes{y}}}
\newcommand{\visz}{\ensuremath{\stokes{z}}}
\newcommand{\visc}{\ensuremath{\stokes{0}}}

\newcommand{\vish}{\ensuremath{\mathcal{V}_H}}
\newcommand{\visv}{\ensuremath{\mathcal{V}_V}}
\newcommand{\visd}{\ensuremath{\mathcal{V}_D}}
\newcommand{\visa}{\ensuremath{\mathcal{V}_A}}
\newcommand{\visr}{\ensuremath{\mathcal{V}_R}}
\newcommand{\visl}{\ensuremath{\mathcal{V}_L}}

\newcommand{\eh}{\ensuremath{|E_H|}}
\newcommand{\ev}{\ensuremath{|E_V|}}
\newcommand{\ed}{\ensuremath{|E_D|}}
\newcommand{\ea}{\ensuremath{|E_A|}}
\newcommand{\er}{\ensuremath{|E_R|}}
\newcommand{\el}{\ensuremath{|E_L|}}

\newcommand{\visopS}[1]{\ensuremath{\hat{\mathcal{S}}_{#1}}}
\newcommand{\visop}[1]{\ensuremath{\hat{\mathcal{V}}_{#1}}}
\newcommand{\visopx}{\ensuremath{\hat{\mathcal{S}}_x}}
\newcommand{\visopy}{\ensuremath{\hat{\mathcal{S}}_y}}
\newcommand{\visopz}{\ensuremath{\hat{\mathcal{S}}_z}}
\newcommand{\visopc}{\ensuremath{\hat{\mathcal{S}}_0}}

\newcommand{\paulx}{\ensuremath{\hat{\sigma}_x}}
\newcommand{\pauly}{\ensuremath{\hat{\sigma}_y}}
\newcommand{\paulz}{\ensuremath{\hat{\sigma}_z}}

\begin{document}

\title{Visibility Stokes parameters as a foundation for \\quantum information science with undetected photons}

\author{Jaroslav Kysela}
\email{kysela.jaroslav@email.cz}
\altaffiliation[Current address: ]{Susi-Nicoletti-Weg 9, Vienna A-1100, Austria}
\affiliation{Faculty of Physics, University of Vienna, Vienna Center for Quantum Science and Technology (VCQ), Vienna, Austria}
\affiliation{Christian Doppler Laboratory for Photonic Quantum Computer, Faculty of Physics, University of Vienna, Vienna, Austria}
\author{Markus Gräfe}
\affiliation{Institute for Applied Physics, Technical University of Darmstadt, Otto-Berndt-Str. 3, 64287 Darmstadt, Germany}
\author{Jorge Fuenzalida}
\email{jorge.fuenzalida@icfo.eu}
\altaffiliation[Current address: ]{ICFO-Institut  de  Ciencies  Fotoniques,  The  Barcelona  Institute  of Science  and  Technology,  08860  Castelldefels  (Barcelona),  Spain}
\affiliation{Institute for Applied Physics, Technical University of Darmstadt, Otto-Berndt-Str. 3, 64287 Darmstadt, Germany}

\begin{abstract}
The framework of measurement operators plays a fundamental role in extracting information about quantum systems. Recently, techniques based on induced coherence have been developed to access the same information for undetected photons. However, there has been a lack of consistent reformulation of quantum operators for these techniques. In this work, we introduce a set of parameters that quantify the polarization of undetected photons based on measured visibilities. Given their similarity to classical counterparts, we refer to them as visibility Stokes parameters. We apply these parameters and the corresponding quantum operators to the problem of quantum state tomography, thoroughly analyzing the environment of undetected photons and its role in the reconstruction process. Because these parameters provide a more intuitive and consistent understanding of the measurement process, we believe that some established quantum information protocols could be adapted for undetected photons.
\end{abstract}

\maketitle

\section{Introduction}

Recent years have witnessed the advent of quantum technologies fueled by quantum information science that uses intrinsic quantum properties of systems such as photons~\cite{nielsen_quantum_2010}. These particles are excellent information carriers and can be transmitted through different media, however, they are absorbed in the measurement process. The development of optical coherence has inspired alternative ways to inspect the state of photons without detecting them through a quantum interference effect called induced coherence without induced emission~\cite{wang1991induced,zou1991induced}. Among these techniques~\cite{hochrainer2022review}, one can find a quantum eraser~\cite{herzog1995eraser,lahiri2017partial}, the quantification of two-photon transverse momenta~\cite{hochrainer2017quantifying}, entanglement certification of a Bell state~\cite{lemos2023one}, and the qubit quantum state tomography~\cite{fuenzalida2022quantum}. The same interference effect has also inspired several applications for probing objects with undetected light~\cite{fuenzalida2024nonlinear}, such as imaging~\cite{lemos_quantum_2014,gilaberte_basset_videorate_2021}, spectroscopy~\cite{kalashnikov2016infrared}, optical coherence tomography~\cite{valles2018optical,paterova2018tunable}, holography~\cite{topfer2022quantum}, and imaging distillation~\cite{fuenzalida2023experimental}. Even though many of these techniques can be counted among the quantum information protocols, a universal recipe that would allow the translation of already established quantum protocols to the context of coherence-based configuration with undetected photons is missing.  

In this work, we make a step towards filling this gap by delineating the relation between the standard quantum state tomography~\cite{james2001tomography} and the tomography of undetected photons~\cite{fuenzalida2022quantum}. More specifically, we focus on the quantum state tomography of undetected qubits represented by the polarization of single photons. We reformulate this technique using a novel type of coherence-based parameters that we term \emph{visibility Stokes parameters} due to their similarity with the Stokes parameters~\cite{stokes1851composition}. But while the latter are based on intensity measurements, the former are extracted from visibilities. The resulting quantum state representation by means of visibility Stokes parameters is strikingly similar to the Bloch vector representation of quantum states in standard quantum tomography. This mutual relation between the two tomography techniques represents a stepping stone for adapting other quantum information techniques to the context of coherence-based quantum operations acting on undetected photons. As a part of our investigation, we present a general analysis of the coherence conditions of the state tomography setup. Moreover, we find how coherence and environment are intertwined when mixed states are considered.

This manuscript is organized as follows: In Sec.~\ref{sec:prelims} we review the standard tomography technique with Stokes parameters, Bloch vectors and Pauli operators. Then, in Sec.~\ref{sec:reconstruct_pure}, we introduce the visibility Stokes parameters that allow us to represent the pure polarization state of the idler photon by visibility measurements performed on the signal photon. The notion of visibility Stokes parameters is generalized for mixed states in Sec.~\ref{sec:mixed}, where a number of complications is discussed and possible remedies are proposed. In Sec.~\ref{sec:algebra} some algebraic properties of the visibility Stokes parameters and their connection to standard Stokes parameters are discussed. In Sec.~\ref{sec:quantum_opers}, we present the quantum operators corresponding to the visibility Stokes parameters, and in Sec.~\ref{sec:postmeas_states}, we calculate the post-measurement states of the idler photon. We conclude in Sec.~\ref{sec:conclusion}.

\section{Preliminaries}
\label{sec:prelims}

The polarization of light and ways to quantify it has a long-standing history in optics~\cite{born2013principles}. One of the first approaches in classical optics was introduced by G. G. Stokes with the so-called Stokes parameters~\cite{stokes1851composition}. Its modern treatment includes the contributions of several authors and concepts such as correlation functions and correlation matrices~\cite{wolf2007introduction}. In the realm of quantum optics, polarization is measured with the help of the Pauli operators~\cite{nielsen_quantum_2010} and employing quantum state tomography~\cite{james2001tomography}. In the following, we review the Stokes parameters and the quantum state tomography techniques in more detail.

\subsection{Stokes parameters}

The Stokes parameters were introduced in 1852~\cite{stokes1851composition} to describe the polarization of classical fields by using four quantities: the ``zeroth'' Stokes parameter $\stokespar{0}$ represents the total intensity of the field, while the parameters $\stokespar{x}$, $\stokespar{y}$, and $\stokespar{z}$ correspond to polarization measurements made in different polarization bases and are defined as
\begin{eqnarray}
    \stokespar{x} = & \ed{^2}-\ea{^2}, \label{eq:stokesclass_x}\\
    \stokespar{y} = & \el{^2}-\er{^2}, \\
    \stokespar{z} = & \eh{^2}-\ev{^2}. \label{eq:stokesclass_z}
\end{eqnarray}
In these formulas, $E_k$ stands for the field's amplitude associated with the polarization mode $k$, where $k$ is either a linear polarization (horizontal $H$, vertical $V$, diagonal $D$, and anti-diagonal $A$) or a circular polarization (left-circular $L$ and right-circular $R$). The total intensity does not depend on the polarization basis, and so
\begin{equation}
        \stokespar{0} = \ed{^2}+\ea{^2}=\el{^2}+\er{^2}=\eh{^2}+\ev{^2}.
\end{equation}

One can remove the dependence of the parameters $\stokespar{x}$, $\stokespar{y}$, and $\stokespar{z}$ on the intensity by dividing them by $\stokespar{0}$. The resulting normalized Stokes parameters allow for a convenient visual representation in the form of a Bloch vector $\vec{r}_I = (x, y, z)$, where
\begin{equation}
    x = \stokespar{x} / \stokespar{0}, \quad y = \stokespar{y} / \stokespar{0}, \quad z = \stokespar{z} / \stokespar{0}.
\end{equation}
The quantum counterpart of these normalized parameters is given by the mean values of Pauli operators \cite{pauli1927quantenmechanik,nielsen_quantum_2010}. These operators read explicitly
\begin{eqnarray}
    \paulx=& \ket{D}\bra{D}-\ket{A}\bra{A}, \label{eq:paulx} \\
    \pauly=&\ket{L}\bra{L}-\ket{R}\bra{R}, \label{eq:pauly} \\
    \paulz=&\ket{H}\bra{H}-\ket{V}\bra{V}. \label{eq:paulz}
\end{eqnarray}
The mean values of Pauli operators for a given quantum state $\rhoi$ coincide with the Bloch vector components as
\begin{equation}
    x = \Tr(\paulx \, \rhoi), \quad y = \Tr(\pauly \, \rhoi), \quad z = \Tr(\paulz \, \rhoi).
\end{equation}
The polarization quantum state can then be reconstructed from the Bloch vector $\vec{r}_I$ using the Bloch representation
\begin{equation}
    \rhoi = \frac{1}{2} (\ident_2 + \vec{r}_I \cdot \vec{\sigma}). \label{eq:bloch_repr}
\end{equation}
The set of all valid polarization states forms a ball referred to as the Bloch sphere, see  Fig.~\ref{fig:explanation}(b). The pure quantum states are located on the surface of the Bloch sphere, while its interior is formed by mixed quantum states with the maximally mixed state in the center. We can parametrize the density matrix of a mixed polarization state in the $H/V$ basis as
\begin{equation}
    \rhoi = \begin{pmatrix}
    \alpha^2 & \alpha \, \beta \, q \, e^{-i \xi} \\ \alpha \, \beta \, q \, e^{i \xi} & \beta^2
    \end{pmatrix},
    \label{eq:mixed_state}
\end{equation}
where $\alpha^2 + \beta^2= 1$, $0 \le q \le 1$, and $\alpha, \beta \ge 0$, $\xi \in [0, 2 \pi)$. The Bloch vector coordinates then read explicitly
\begin{eqnarray}
    x & = & 2 \, \alpha \, \beta \, q \, \cos(\xi), \label{eq:bloch_coord_x} \\
    y & = & 2 \, \alpha \, \beta \, q \, \sin(\xi), \label{eq:bloch_coord_y} \\
    z & = & \alpha^2 - \beta^2. \label{eq:bloch_coord_z}
\end{eqnarray}

\subsection{Quantum state tomography}

\begin{figure}[htpb]
    \centering
    \includegraphics[width=\linewidth]{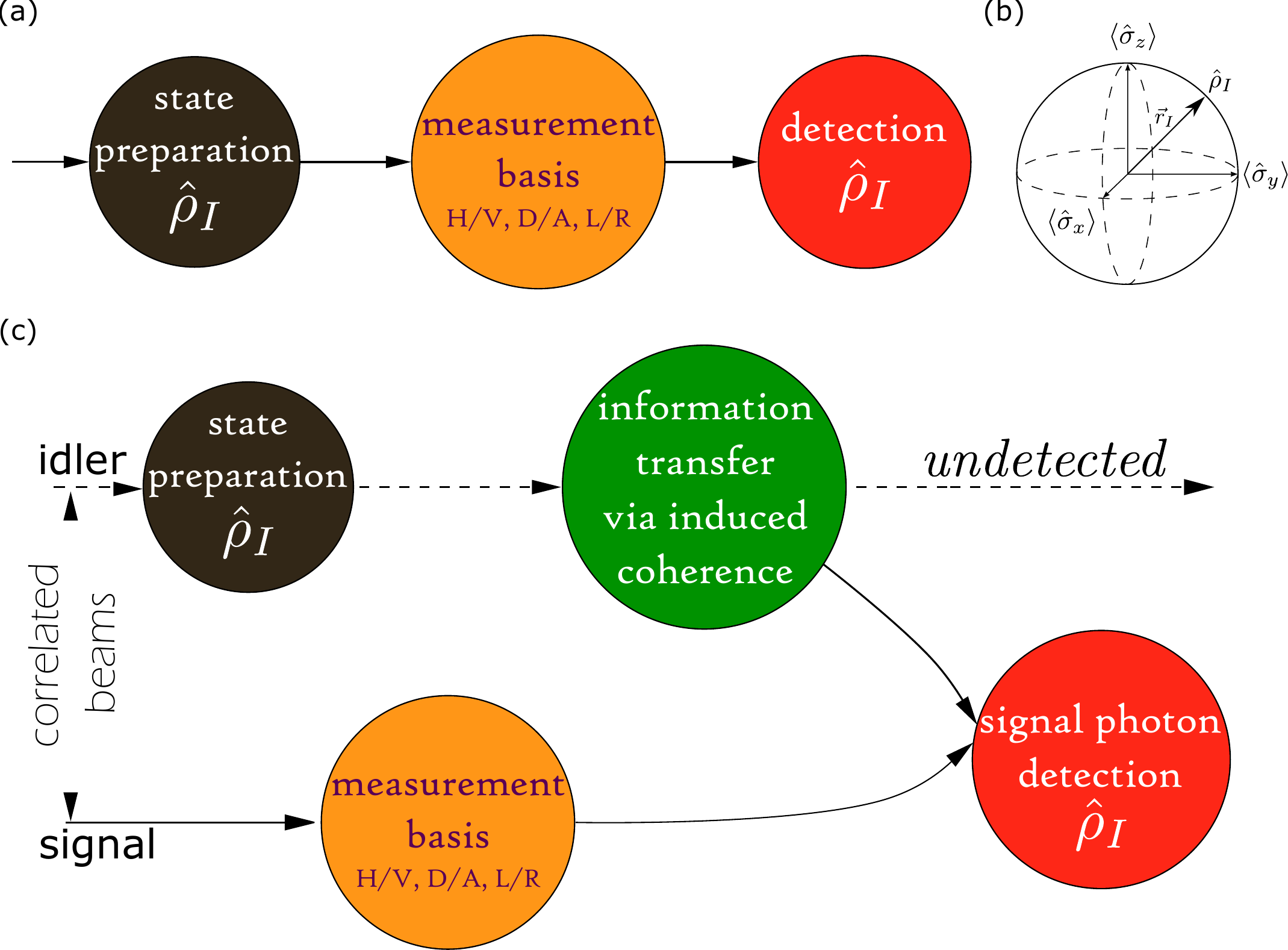}
    \caption{\textbf{Quantum state tomography of photonic qubits.} (a) An ensemble prepared in an unknown state $\rhoi$ can be reconstructed by measurements in different bases. This technique is known as the quantum state tomography. The state $\rhoi$ can be visualized as a point in the Bloch sphere depicted in (b). (c) The idler (top dashed line) and the signal (bottom solid line) photons form a photon pair (that can be produced via SPDC) and are employed for quantum state tomography of undetected photons. The idler is prepared in an initial state $\rhoi$, which is transferred to the signal through induced coherence. The idler photon remains undetected. Single-photon interference patterns shown by the signal allow us to reconstruct the idler state $\rhoi$.}
    \label{fig:explanation}
\end{figure}

The method that allows to determine the quantum state $\rhoi$ of a quantum system such as a photon is called quantum state tomography~\cite{james2001tomography,altepeter2005photonic}. In the standard quantum tomography, one has an ensemble of identical systems in state $\rhoi$ and subjects each system to a measurement in a certain basis, cf. Fig.~\ref{fig:explanation}(a). The measurement results provide sufficient information to reconstruct the state $\rhoi$. For qubits, the measurement bases are usually given as the eigenbases of Pauli operators, and the state can be visually represented in the Bloch sphere depicted in Fig.~\ref{fig:explanation}(b). If the ensemble is composed of photons, the measurement typically leads to the photons' destruction.

However, the state reconstruction can also be carried out non-destructively. The quantum state tomography of undetected photons~\cite{fuenzalida2022quantum}, depicted in Fig.~\ref{fig:explanation}(c), uses a pair of photons, for convenience referred to as the signal and the idler, which can be generated in a spontaneous parametric down-conversion process (SPDC). In this technique, the state $\rhoi$ of the idler photon is reconstructed by measuring its partner, the signal photon, while the idler photon remains undetected. The information about the quantum state is transferred from the idler to the signal via an induced coherence configuration. The state $\rhoi$ is then reconstructed from the interference patterns measured for the signal photon.

For future reference and to emphasize the differences between the present work and Ref.~\cite{fuenzalida2022quantum}, we list below the individual steps that constitute the quantum state tomography of polarization of undetected photons:
\begin{enumerate}
    \item A non-linear interferometer (for details see Sec.~\ref{sec:reconstruct_pure}) with two sources of photons is employed to produce a pair of photons, the signal and the idler. The effect of induced coherence turns the state of these photons into a superposition of a reference state and the state $\hat{\rho}_I$ that is to be reconstructed. This step, which is identical for both Ref.~\cite{fuenzalida2022quantum} and our work, is schematically represented by the two upper bubbles in Fig.~\ref{fig:explanation}(c). Unlike in Ref.~\cite{fuenzalida2022quantum}, our approach considers the most general coherence conditions in the experimental setup.
    \item An identical ensemble of photon pairs is produced by the interferometer and subjected to several rounds of measurements. In each round, the signal photons' state is locally adjusted to a particular form and the signal photons are subjected to measurements in a given polarization basis. This step is schematically represented by the two lower bubbles in Fig.~\ref{fig:explanation}(c). While in Ref.~\cite{fuenzalida2022quantum} the measurement consists of two rounds and the signal photons are always measured in $H/V$ basis, in our work the measurement comprises three rounds and in each round the signal photons are measured in one of Pauli eigenbases $H/V$, $D/A$, and $L/R$.  
    \item For each round, the phase between the two photon sources is varied and the interference pattern in the rate of signal photons is recorded. The idler photons remain undetected and leave the interferometer, while the interference patterns for signal photon detections are post-processed and relevant quantities are extracted. In Ref.~\cite{fuenzalida2022quantum}, the visibilities and the relative phase shift between two interference patterns were used to reconstruct the polarization state $\hat{\rho}_I$. In our work, we instead need only the visibilities of the three interference patterns. This way, the dependence of the reconstructed state on the experimental phase backlash is removed.
\end{enumerate}

Apart from the technical improvements listed above, there is a more conceptual and notable difference --- we introduce visibility Stokes parameters that represent a convenient tool for analyzing quantum measurements based on coherence. These new parameters are determined by the visibilities of the recorded interference patterns and can be seen as counterparts of standard Stokes parameters that are based on intensities. The visibility Stokes parameters can be visualized in the form of the visibility Bloch vector and we show its connection to the standard Bloch representation of mixed states. In the upcoming sections we describe our technique in detail and perform a thorough theoretical analysis, first for pure states and then for mixed states.

\section{Pure states}
\label{sec:reconstruct_pure}

% \blue{Expand explanation ---comment 5}

\begin{figure*}[ht]
\centering
\includegraphics[width=\linewidth]{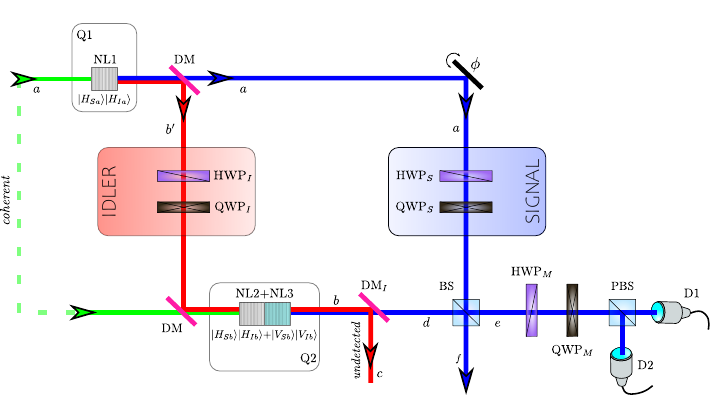}
\caption{\textbf{Imbalanced Zou-Wang-Mandel interferometer}. The pump, the idler photon, and the signal photon are represented by the green, red, and blue lines, respectively. Each photon pair is created in a coherent superposition of three crystals (NL1, NL2, and NL3). Signal beams emitted from each crystal are superposed in a beam splitter (BS), and those that leave the upper port are then split into $H$ and $V$ polarization by a polarizing beam splitter (PBS) and detected by detectors D1 and D2. Idler beams are overlapped to erase the which-source information, thus, inducing coherence on the signal photon. Consequently, the intensities of the signal photon recorded by the detectors D1 and D2 exhibit interference as the independent phase $\phi$ is varied. The introduction of an additional quarter-wave plate QWP$_S$ in the signal photon path together with half- and quarter-waveplates HWP$_M$ and QWP$_M$ in the detection path allows one to estimate an arbitrary polarization state of the idler photon. See the text for details. DM stands for a dichroic mirror, and HWP$_I$ and QWP$_I$ stand for half- and quarter-waveplates in the path of the idler photon, respectively.}
\label{fig:setup}
\end{figure*}

We begin our discussion by considering pure quantum states of the idler photon. The entire tomography procedure can schematically be described as follows:
\begin{enumerate}
    \item Idler initial state --- The setup shown in Fig.~\ref{fig:setup} is utilized, where the idler photon with the unknown pure state $\ket{\psi_I}$ propagates along (red) path $b'$.
    \item Signal bases and detection --- Three rounds of measurements are performed on an ensemble of identically prepared signal photons. For each round, the signal photon's state is set to a particular form (see blue box in Fig.~\ref{fig:setup}) and the phase $\phi$ in path $a$ is scanned while detection events are recorded by detectors D1 and D2.
    \item Idler state reconstruction --- The visibility of the resulting interference pattern for each of the three rounds is determined. The visibility Stokes parameters are calculated from these visibilities, which parameters contain enough information to reconstruct the pure state $\ket{\psi_I}$.
\end{enumerate}
In the following, we describe each of these three steps in a separate subsection in more detail.

\subsection{Measurement setup}
\label{sec:pure_meas}

The physical setup of our technique is identical to the one used for the tomography of undetected photons in Ref.~\cite{fuenzalida2022quantum} and is formed by the imbalanced Zou-Wang-Mandel (ZWM) interferometer. This interferometer, depicted in Fig.~\ref{fig:setup}, consists of two sources of photon pairs, Q1 and Q2. The source Q1 produces photon pairs in a separable state $\ket{\psi_S} \otimes \ket{\psi_{I}}$, where $\ket{\psi_{I}}$ is the polarization state of the idler photon that is unknown to us and that we wish to reconstruct
\begin{eqnarray}
    \ket{\psi_{I}} = \alpha \ket{H_{Ib'}} + \beta e^{i \xi} \ket{V_{Ib'}},
    \label{eq:idler_state}
\end{eqnarray}
where $\ket{k_{Ib'}}$ denotes a one-photon state of the idler photon carrying a polarization $k$ and traveling through the path $b'$, and where $\alpha^2 + \beta^2 = 1$, $\alpha, \beta \ge 0$, and $\xi \in [0, 2 \pi)$. The state of the signal photon $\ket{\psi_S}$ is chosen depending on the reconstruction technique and, generally, can be prepared in any pure state by including retarder plates on its path. Here, a half-wave plate HWP$_S$ and a quarter-wave plate QWP$_S$ are used so that the signal photon is set to
\begin{equation}
    \ket{\psi_{Sa}} = \frac{1}{\sqrt{2}}( \ket{H_{Sa}} + e^{i \zeta} \ket{V_{Sa}}),
    \label{eq:signal_state}
\end{equation}
with $\zeta \in [0, 2\pi)$.

The second source Q2 produces a maximally entangled state $\ket{\Psi}$ of two photons propagating along path $b$, where $\ket{\Psi} = (1/\sqrt{2})(\ket{H_{Sb}}\otimes\ket{H_{Ib}}+\ket{V_{Sb}} \otimes \ket{V_{Ib}})$. Due to the overlap of idler beams coming from Q1 and Q2, path $b'$ becomes $b$. This is a crucial step towards erasing the which-source information. The second step is recombining the signal beams in a beam splitter. The entire experimental setup produces a \textit{single} photon pair at a time, whose state is a coherent superposition of state $\ket{\psi_{S}}\otimes\ket{\psi_{I}}$ coming from the source Q1 and $\ket{\Psi}$ coming from the source Q2. The relative pump power for sources Q1 and Q2 is chosen such that the total state of the quantum system just before the beam splitter in a lossless scenario reads
\begin{multline}
    \ket{\psi_{d}} \propto (\ket{H_{Sb}}\ket{H_{Ic}}+\ket{V_{Sb}}\ket{V_{Ic}}) \\
    + e^{i \phi} (\ket{H_{Sa}}+e^{i \zeta} \ket{V_{Sa}})(\alpha \ket{H_{Ic}}+\beta e^{i \xi} \ket{V_{Ic}}). \nonumber
\end{multline}
The relevant part of the state after the beam splitter is given by
\begin{eqnarray}
        \ket{\psi_{e}} & \propto & \ket{H_{Sb}}\Bigg( \left(1 + \alpha e^{i \phi} \right)\ket{H_{Ic}}+\beta e^{i (\phi + \xi)} \ket{V_{Ic}}\Bigg) 
        \label{eq:expr} \\ %\nonumber \\
    & + & \ket{V_{Sb}} \Bigg( \alpha e^{i (\phi + \zeta)} \ket{H_{Ic}}+ \left(1 + \beta e^{i (\phi+\xi+\zeta)} \right)\ket{V_{Ic}} \Bigg). \nonumber
\end{eqnarray}

\subsection{Interference measurements}
\label{sec:pure_interf}

During the actual tomography measurement, the setup is kept fixed except for the interferometric phase $\phi$. The two detectors D1 and D2 thus record single-photon interference patterns, whose forms can be derived from the expression in Eq.~\eqref{eq:expr}. The visibility $\mathcal{V}$ of the interference pattern is given by
\begin{equation}
\mathcal{V}=\frac{\mathcal{I}_{\textup{max}}-\mathcal{I}_{\textup{min}}}{\mathcal{I}_{\textup{max}}+\mathcal{I}_{\textup{min}}},
\end{equation}
where $\mathcal{I}_{\textup{max}}$ ($\mathcal{I}_{\textup{min}}$) stands for the intensity maximum (minimum). The signal photons can be detected in any polarization basis. For a fixed value of the relative phase $\zeta$ in the signal photon's state in Eq.~\eqref{eq:signal_state} one can determine the visibilities in bases $H/V$, $D/A$, and $L/R$. These turn out to be
\begin{equation}
    \vish = \alpha, \quad \visv = \beta,
    \label{eq:vis_hv}
\end{equation}\begin{equation}
    \vis{D/A} = \frac{2\sqrt{1\pm \cos(\zeta)}}{2\pm \cos(\zeta)} \, \sqrt{\frac{1\pm2\alpha\beta\cos(\xi)}{2}},
    \label{eq:vis_da}
\end{equation}
\begin{equation}
    \vis{L/R} = \frac{2\sqrt{1 \mp \sin(\zeta)}}{2 \mp \sin(\zeta)} \, \sqrt{\frac{1 \pm 2\alpha\beta\sin(\xi)}{2}}.
    \label{eq:vis_rl}
\end{equation}
Visibilities in bases $D/A$ and $L/R$ depend on the choice of relative phase $\zeta$. There does not exist $\zeta$ such that the prefactors in expressions \eqref{eq:vis_da} and \eqref{eq:vis_rl} are all equal, but we can set $\zeta = \pi/2$ and $\zeta = 0$ for the $D/A$ and $L/R$ bases, respectively, to obtain
\begin{equation}
    \vis{D/A} = \sqrt{\frac{1\pm2\alpha\beta\cos(\xi)}{2}},
    \label{eq:vis_da_spec}
\end{equation}
and
\begin{equation}
    \vis{L/R} = \sqrt{\frac{1 \pm 2\alpha\beta\sin(\xi)}{2}}.
    \label{eq:vis_rl_spec}
\end{equation}
In Appendix \ref{sec:vispure} the general formulas for the visibilities in the three bases are shown for arbitrary pumping powers, transmission coefficients, and signal photon states.

\subsection{Visibility Stokes parameters}

In the previous step, the explicit formulas for the visibilities in three bases $H/V$, $D/A$, and $L/R$ were found. From Eqs.~\eqref{eq:vis_hv}, \eqref{eq:vis_da_spec}, and \eqref{eq:vis_rl_spec} it is straightforward to check that
\begin{eqnarray}
    \visd^2-\visa^2 & = & 2 \, \alpha\beta\cos(\xi), \label{eq:stokes_x} \\
    \visl^2-\visr^2 & = & 2 \, \alpha\beta\sin(\xi), \label{eq:stokes_y} \\
    \vish^2-\visv^2 & = & \alpha^2 - \beta^2, \label{eq:stokes_z}
\end{eqnarray}
and
\begin{equation}
    \visd^2+\visa^2 = 1, \quad 
    \visl^2+\visr^2 = 1, \quad
    \vish^2+\visv^2 = 1. \label{eq:sum_pure}
\end{equation}

These expressions are in the exact correspondence with Stokes parameters in Eqs.~\eqref{eq:stokesclass_x}--\eqref{eq:stokesclass_z} if we identify
\begin{eqnarray}
    \visx & \equiv & \visd^2 - \visa^2, \label{eq:s_x} \\
    \visy & \equiv & \visl^2 - \visr^2, \label{eq:s_y} \\
    \visz & \equiv & \vish^2 - \visv^2. \label{eq:s_z}
\end{eqnarray}
Note that for these \emph{visibility} Stokes parameters $\visx$, $\visy$, and $\visz$ we use a calligraphic font to differentiate them from the standard Stokes parameters $\stokespar{x}$, $\stokespar{y}$, and $\stokespar{z}$. This distinction will prove important later on when discussing the mixed states. The three visibility Stokes parameters allow us to calculate the coefficients $\alpha$, $\beta$, and the phase $\xi$ of the unknown state in Eq.~\eqref{eq:idler_state} purely from visibility measurements in different bases.

For the case of pure idler states $\ket{\psi_I}$, the three parameters $\visx$, $\visy$, and $\visz$ coincide with the three coordinates $(x, y, z) = \vec{r}_I$ of the Bloch vector associated with the idler photon's state. We thus recover the Bloch representation of the idler's polarization state in Eq.~\eqref{eq:bloch_repr}. This demonstrates the correspondence between the traditional direct polarization measurement and our visibility measurement of undetected photons.

Let us note that in the discussion above, we used the concept of mutually unbiased bases~\cite{durt_mutually_2010}. Namely, we can measure visibilities for both modes of basis $B$ by setting the state of the signal photon such that it belongs to a basis $B'$ that is unbiased to $B$. For example, to measure simultaneously both visibilities $\alpha$ and $\beta$ in $H/V$ basis, see Eq.~\eqref{eq:vis_hv}, we have to set the signal photon into the state in Eq.~\eqref{eq:signal_state}, which is mutually unbiased with $\ket{H_{Sa}}$ and $\ket{V_{Sa}}$.

\section{Mixed states}
\label{sec:mixed}

So far, we have considered only pure polarization states of the idler photon. For our tomography technique to be applicable in a practical setting, we present in this section its generalization for mixed states. We again proceed in a stepwise manner. The procedure is essentially identical to that for pure states with the exception that the theoretical treatment is more involved and there are extra caveats that have to be considered:
\begin{enumerate}
    \item Idler initial state --- The idler photon's state is now a mixed state $\hat{\rho}_I$. In Sec.~\ref{sec:model} we show how we model this state using a larger pure state. The evolution of both signal and idler photons through the measurement setup is studied in Sec.~\ref{sec:evol}.
    \item Signal bases and detection --- The analytical formulas for detection probabilities in different measurement bases of the signal photon are derived in Sec.~\ref{sec:evol}.
    \item Idler state reconstruction --- The form of visibilities is found in Sec.~\ref{sec:visib}. The visibility Stokes parameters for mixed states are defined in Sec.~\ref{sec:stokes} with the aim of being as close as possible to the standard Stokes parameters.
    \item Role of asymmetric coherence --- We observe that when the environmental conditions are completely unknown, the uncertainty in the reconstruction of mixed states increases with the decoherence in the systems. We discuss what additional constraints have to be adopted to enable the mixed state reconstruction in Sec.~\ref{sec:coherence}.
\end{enumerate}
Each of these steps is discussed in detail below.

\subsection{Modelling mixed states with pure states}
\label{sec:model}

In order to treat mixed idler states in our tomography setup we need to represent mixed states in the framework of induced coherence. We do so by modelling mixed states with pure states as explained below. Mixed states can be represented as pure states living in a larger Hilbert space using a mathematical trick known as \emph{purification} \cite{nielsen_quantum_2010}, where a given mixed state $\hat{\rho} = \sum_j p_j \ket{\varphi_j}\bra{\varphi_j}$ is associated with a pure state $\ket{\psi} = \sum_j \sqrt{p_j} \ket{\varphi_j} \ket{a_j}$, while $\{ \ket{a_j} \}_j$ form an orthonormal basis of an extra ancillary Hilbert space. The partial trace over this extra space then gives the original mixed state as $\hat{\rho} = \mathrm{Tr}_A (\ket{\psi} \bra{\psi})$. The choice of $\{ \ket{a_j} \}_j$ is arbitrary, and any orthonormal basis will do. In the following, we adopt a more physics-inspired approach that bears some similarity to the purification yet shows important differences.

Let us assume without loss of generality that the mixed state $\rhoi$ of the idler photon results from some decoherence processes in the state-preparation stage, which processes act unitarily on the composite pure state $(\alpha_0 \ket{H} + \beta_0 \ket{V}) \otimes \ket{e}$. The vector $\ket{e}$ denotes the pure state of all the other degrees of freedom of a photon, which we from now on collectively refer to as the \emph{environment}. The decoherence processes might be polarization-dependent, and so the final state of the idler photon just before entering Q2 attains the form
\begin{equation}
    \ket{\psi_I} = \alpha \ket{H_{b'}} \otimes \ket{e_H} + \beta e^{i \, \xi} \ket{V_{b'}} \otimes \ket{e_V},
    \label{eq:larger_idler_pure}
\end{equation}
where $\ket{e_H}$ and $\ket{e_V}$ are the final states of the environment for the $\ket{H}$ and $\ket{V}$ idler photon polarization mode, respectively. These states are not in general orthogonal and without loss of generality we can assume that $0 \leq \braket{e_H}{e_V} \leq 1$. If we make the following identification
\begin{equation}
    q = \braket{e_H}{e_V}
    \label{eq:overlap}
\end{equation}
it is straightforward to verify that the partial trace over the environment returns the desired mixed state
\begin{eqnarray}
\rhoi & = & \Tr_{e}(\ket{\psi_I}\bra{\psi_I}).
\end{eqnarray}
The parameter $q$ in the parametrization of $\rhoi$ in Eq.~\eqref{eq:mixed_state} quantifies the degree of coherence of the state in the $H/V$ basis: for $q = 1$, the state is pure, and for $q = 0$ it is completely dephased (in the $H/V$ basis). Note that, unlike the standard purification procedure, here, the states $\ket{e_j}$ are manifestly not orthogonal. However, there is still large freedom in their choice as $q = \braket{e_H}{e_V} = \bra{e_H} U^\dagger U \ket{e_V}$ for any unitary $U$. We can thus model a mixed state $\rhoi$ by $\ket{\psi_I}$ in Eq.~\eqref{eq:larger_idler_pure} as long as the overlap of environmental states satisfies Eq.~\eqref{eq:overlap}. Analogously, the composite state of the second source is now
\begin{equation}
    \ket{\Psi} = \frac{1}{\sqrt{2}}(\ket{H_b, H_b} + \ket{V_b, V_b}) \otimes \ket{e_\Psi}. \label{eq:source_2_psi}
\end{equation}
We assume that the $H$- and $V$-polarized terms in Bell state $\ket{\Psi}$ are perfectly coherent and so they share the same environmental state $\ket{e_\Psi}$.

\subsection{Photon evolution through setup}
\label{sec:evol}

In what follows, we present on a more abstract level the generalization of calculations done for pure states in Sec.~\ref{sec:pure_meas}. The results form the stepping stone for discussing mixed state reconstruction, visibilities, and Stokes parameters later on. To account for losses in the setup, the imperfect transmission of objects encountered by the idler photon is modeled as a beam splitter in path $b'$ with the reflection coefficient equal to $\sqrt{1-T^2}$ that reflects the idler photon into path $w$.\footnote{Strictly speaking, $T$ can be polarization-dependent as discussed in Ref.~\cite{fuenzalida2022quantum}. However, here we consider the case of homogeneous losses for every polarization component.} The corresponding term we denote by $\ket{\psi_w}$. By $\ket{\psi_S}$ we denote the state of the signal photon created in Q1 and propagating along path $a$ after it traversed HWP$_S$ and QWP$_S$, i.e., just before it enters the beam splitter BS. Similarly, by $\ket{\psi_I}$ we denote the state of the idler photon created in Q1 and propagating along path $b'$ after it traversed HWP$_I$ and QWP$_I$, which is now of the form of Eq.~\eqref{eq:larger_idler_pure}. This state propagates through the source Q2 unaffected and is reflected by the dichroic mirror DM$_I$ out of the setup. The state of signal and idler photons just before the projective measurement reads
\begin{multline}
    \ket{\psi} = \mathcal{N} \cdot \mathrm{BS} \cdot \mathrm{DM}_{I} \cdot (T e^{i \phi} \ket{\psi_{S}}\otimes\ket{\psi_{I}} \,  \\ + \sqrt{1-T^2} \, e^{i \phi} \ket{\psi_{S}}\otimes\ket{\psi_{w}} + \rp \, \ket{\Psi}), \label{eq:initial_state}
\end{multline}
where $\phi$ is the relative phase between the two sources, $\rp$ determines the relative pump power between the first and the second source as $\rp^2$, and $\mathcal{N}$ is a normalization factor equal to $\mathcal{N} = 1/\sqrt{1+\rp^2}$.

Note that the state of the idler photon has two terms. One term is $\ket{\psi_I}$ coming from Q1, and the other term is included in $\ket{\Psi}$ coming from Q2. Both terms are reflected off a dichroic mirror DM$_I$ into path $c$ out of the setup. The dichroic mirror DM$_I$ acts only on the idler state, and the beam splitter BS acts only on the signal state. We can, therefore, rewrite the above state into
\begin{multline}
    \ket{\psi} = \mathcal{N} \, \big( T e^{i \phi} \ket{\psi'_{S}}\otimes\ket{\psi'_{I}} \\ + \sqrt{1-T^2} \, e^{i \phi} \ket{\psi'_{S}}\otimes\ket{\psi_{w}} + \rp \, \ket{\Psi'} \big),
    \label{eq:superpos}
\end{multline}
where we defined $\ket{\psi'_{S}} = \mathrm{BS} \cdot \ket{\psi_{S}}$, $\ket{\psi'_{I}} = \mathrm{DM}_{I} \cdot \ket{\psi_{I}}$ and $\ket{\Psi'} = \mathrm{BS} \cdot \mathrm{DM}_{I} \cdot \ket{\Psi}$. Remember that $\braket{\psi_w}{\psi'_I} = \braket{\psi_w}{\Psi'} = 0$.

At this point, we subject the signal photon to projective measurements, embodied by projectors $\hat{\Pi}_k$ and $\hat{\Pi}_k^\perp = \ident - \hat{\Pi}_k$. For a given orthonormal basis $\{ \ket{k}, \ket{k^\perp}\}$ the two projectors are given by $\hat{\Pi}_k = \ket{k_S} \bra{k_S} \otimes \ident$ and $\hat{\Pi}_k^\perp = \ket{k_S^\perp} \bra{k_S^\perp} \otimes \ident$, where the identity acts on the idler photon as well as the environment. The probability of measuring $\hat{\Pi}_k$ for state in Eq.~\eqref{eq:superpos} is easily shown to be
\begin{equation}
    \bra{\psi} \hat{\Pi}_k \, \ket{\psi} = \mathcal{N}^2 (c_k + 2 \, \rp \, T \, \mathrm{Re}(e^{-i \phi} z_k)),
    \label{eq:interference}
\end{equation}
where we defined numerical quantities
\begin{eqnarray}
    c_k & = & \bra{\psi'_{S}} \hat{\Pi}_k \ket{\psi'_{S}} + \rp^2 \bra{\Psi'} \hat{\Pi}_k \ket{\Psi'}, \label{eq:c_j} \\
    z_k & = & \bra{\psi'_{S}}\otimes\bra{\psi'_{I}} \hat{\Pi}_k \ket{\Psi'}. \label{eq:z_j}
\end{eqnarray}
Before we proceed to discuss the visibilities of these detection probabilities, let us emphasize that there is not a single term in Eq.~\eqref{eq:interference} that would contain inner products of $\ket{\psi'_{I}}$ with itself or its projections onto some subspace. For that reason, there is no term that would contain the inner product $\braket{e_H}{e_V}$ \eqref{eq:overlap}, whose determination is necessary for the characterization of the mixed state $\rhoi$. This observation is independent of particular forms of $\ket{\psi'_{S}}$ and $\ket{\Psi'}$ and is thus quite general. We, therefore, conclude with the important statement that the imbalanced ZWM interferometer in Fig.~\ref{fig:setup} does \emph{not} allow for the perfect reconstruction of mixed states unless additional constraints are enforced. We come back to this problem in Sec.~\ref{sec:coherence}.

\subsection{Visibilities}
\label{sec:visib}

In this section, we carry out the derivation of visibilities in different bases in a way analogous to that in Sec.~\ref{sec:pure_interf}. When the signal photon's polarization is measured in an orthonormal basis $\{ \ket{k}, \ket{k^\perp}\}$, the probability of detecting $\ket{k}$ is given in Eq.~\eqref{eq:interference}, which profile exhibits interference as the phase $\phi$ is varied. The visibility of the emergent interference pattern is given by (see Appendix~\ref{sec:vismixed})
\begin{equation}
    \vis{k} = 2 \, \rp \, T \, \frac{|z_k|}{c_k},
    \label{eq:vis_j}
\end{equation}
where $c_k$ and $z_k$ are defined in Eqs.~\eqref{eq:c_j} and \eqref{eq:z_j}, respectively. This expression attains a conveniently simple form when we set $\rp = 1$ and choose the signal photon's state $\ket{\psi_{S}}$ to be unbiased with both $\ket{k}$ and $\ket{k^\perp}$, i.e., $|\braket{\psi_S}{k}| = |\braket{\psi_S}{k^\perp}| = 1/\sqrt{2}$. It reduces to (see Appendix~\ref{sec:vis_mub})
\begin{equation}
    \vis{k} = T |\braket{\psi'_I}{k^\ast}\ket{e_\Psi}|,
    \label{eq:visk_simple}
\end{equation}
where $\ket{k^\ast}$ is the ``complex conjugate'' of vector $\ket{k}$ in the sense that if $\ket{k} = U \ket{H}$ for some unitary $U$, then $\ket{k^\ast} = U^\ast \ket{H}$, where the star stands for the complex conjugation. The visibilities are thus equal to the overlap between the idler state (with its environment) and the complex conjugate basis state (with environment $\ket{e_\Psi}$). 

For completeness, let us list the visibilities of detection probabilities for the typical choice of single-qubit bases $H/V$, $D/A$, and $L/R$. For these, we get $(k,k^\ast) \in \{(H,H), (V,V), (D,D), (A,A), (L,R), (R,L)\}$ and the visibilities explicitly read
\begin{eqnarray}
    \vish & = & T \, \alpha \, |\braket{e_H}{e_\Psi}|, \quad \visv = T \, \beta \, |\braket{e_V}{e_\Psi}|, \label{eq:vishv_form} \\
    \vis{D / A} & = & (T/\sqrt{2}) |\alpha \braket{e_H}{e_\Psi} \pm \beta e^{-i \xi} \braket{e_V}{e_\Psi}|, \label{eq:visda_form} \\
    \vis{L / R} & = & (T/\sqrt{2}) |\alpha \braket{e_H}{e_\Psi} \pm i \beta e^{-i \xi} \braket{e_V}{e_\Psi}|. \label{eq:vislr_form}
\end{eqnarray}
Compare these expressions with Eqs.~\eqref{eq:vis_hv}--\eqref{eq:vis_rl}. The products of environmental states apparently play an important role and for convenience we introduce the following notation
\begin{align}
    m_H & = |\braket{e_H}{e_\Psi}|, \quad m_V = |\braket{e_V}{e_\Psi}|, \label{eq:mHmV} \\
    \Delta \varphi & = \arg \left(\braket{e_H}{e_\Psi} \right) - \arg \left(\braket{e_V}{e_\Psi} \right). 
\end{align}
These parameters quantify the coherence of the $H$ and $V$ modes of the idler state with the second source Q2 and their properties are further studied in Sec.~\ref{sec:coherence}.

\subsection{Visibility Stokes parameters for mixed states}
\label{sec:stokes}

In the previous section, formulas \eqref{eq:vishv_form}--\eqref{eq:vislr_form} connect the form of environmental states and the visibilities of signal photon detection rates in individual Pauli bases. It is easy to verify from these expressions that
\begin{equation}
    \vish^2 + \visv^2 = \visd^2 + \visa^2 = \visl^2 + \visr^2 = \visc,
    \label{eq:visc_0}
\end{equation}
where we defined
\begin{equation}
    \visc \equiv T^2(\alpha^2 m_H^2 + \beta^2 m_V^2).
    \label{eq:visc}
\end{equation}
The sum of squares of visibilities in the two orthogonal states is thus constant. This constant quantifies the level of coherence between the first and the second source, and we refer to it as the \emph{zeroth visibility Stokes parameter} $\visc$. It is easy to check that $0 \leq \visc \leq T^2 \leq 1$ (compare with Eq.~\eqref{eq:sum_pure}). In Appendix~\ref{sec:visc_relations} we show that $\vis{k}^2 + \vis{k^\perp}^2 = \visc$ for any orthonormal basis $\{ \ket{k}, \ket{k^\perp}\}$, not just the Pauli bases.

We can define the other visibility Stokes parameters for mixed states in the exact same way as we did for pure states in Eqs.~\eqref{eq:s_x}--\eqref{eq:s_z} except that now the visibilities are given by Eq.~\eqref{eq:visk_simple}. The visibility Stokes parameters then explicitly read
\begin{eqnarray}
    \visx & = & T^2 \, 2 \, \alpha \, \beta \, m_H \, m_V \, \cos(\xi + \Delta \varphi), \label{eq:vis_da_mixed} \\
    \visy & = & T^2 \, 2 \, \alpha \, \beta \, m_H \, m_V \, \sin(\xi + \Delta \varphi), \label{eq:vis_lr_mixed} \\
    \visz & = & T^2 \, (\alpha^2 \, m^2_H - \beta^2 \, m^2_V). \label{eq:vis_hv_mixed}
\end{eqnarray}
The transmission coefficient $T$ can be measured for a given experimental setup independently of a particular state of the idler photon, and one can thus adjust the formulas above by removing $T$. For this reason, from now on, we effectively set $T = 1$.

There are a number of problems associated with these expressions. First, the coherence terms $m_H$ and $m_V$ always appear in the product with $\alpha$ and $\beta$, respectively, which precludes the determination of the values of $\alpha$ and $\beta$ alone. Second, as discussed in Sec.~\ref{sec:model}, these formulas explicitly depend on the environment of the second source $\ket{e_\Psi}$, and there is no quantity $q$ present in the formulas. These two issues together imply that a given triple of visibility Stokes parameters $(\visx, \visy, \visz)$ is consistent with many states in the standard Bloch sphere. Moreover, the environment can have many degrees of freedom, and it can happen that states $\ket{e_H}$ and $\ket{e_V}$ turn out to be orthogonal to $\ket{e_\Psi}$, in which case all the visibility parameters are zero: $\visx = \visy = \visz = 0$. We would then misinterpret our results as corresponding to the maximally mixed state, whose Bloch vector is a zero vector.

In the next section, the aforementioned issues are discussed in detail, and possible solutions to the ambiguity of the visibility Stokes parameters are proposed.

\subsection{Role of asymmetric coherence}
\label{sec:coherence}

\begin{figure}
    \centering
    \includegraphics[width=\linewidth]{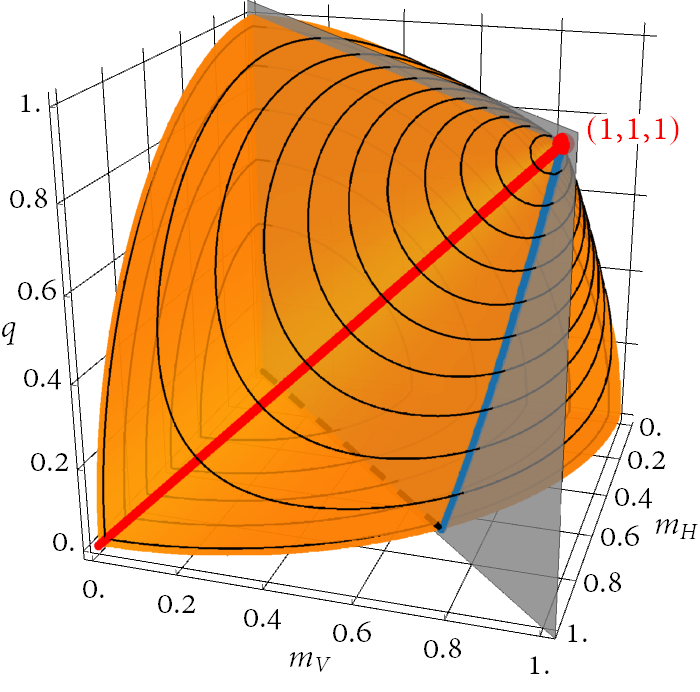}
    \caption{\textbf{Region plot of valid values of coherence parameters when $\Delta \varphi = 0$.} The orange solid corresponds to values that satisfy inequality \eqref{eq:coherence_inequality}. The black contours mark the equipotential curves with $q^2 + m_H^2 + m_V^2 = const.$, and the gray plane marks the constraint $m_H = m_V$. The role of all three parameters is symmetric. As the value of any of the parameters gets close to one, the region of allowed values for the other two gets smaller. The extreme point, depicted as a red dot, is when all coherence parameters attain the value of one. The red line corresponds to the allowed values of $q$ and $m_V$ when $m_H$ is set to 1. The blue line corresponds to the case when $m_H = m_V = \sqrt{(1+q)/2}$. The dashed dim line corresponds to $m_H = m_V \leq 1/\sqrt{2}$, for which the parameter $q$ can take on any value between 0 and 1. For details, see the main text.}
    \label{fig:cohplot}
\end{figure}

The quantities $q$, $m_H$, $m_V$, and $\Delta \varphi$ defined in the previous sections are not completely independent of each other. As shown in Appendix~\ref{sec:constraint}, they have to satisfy the inequality
\begin{equation}
    1-\left(q^2+m_H^2+m_V^2\right) + 2 \, q \, m_H \, m_V \cos(\Delta \varphi) \geq 0,
    \label{eq:coherence_inequality}
\end{equation}
where additionally $0 \leq q \leq 1$, $0 \leq m_H \leq 1$, and $0 \leq m_V \leq 1$. The inequality \eqref{eq:coherence_inequality} embodies a certain type of transitivity of coherence: when any two of the coherence parameters are close to unity, say $m_H$ and $m_V$, the third parameter, $q$, has to be almost unity as well. In terms of polarization modes and the physical setup in Fig.~\ref{fig:setup}: when $H$ polarization mode from Q1 is highly coherent with $H$ produced by Q2 and also $V$ mode from Q1 is highly coherent with $V$ produced by Q2, then due to the fact that NL2 and NL3 in Q2 are highly coherent with each other, also the two polarization modes $H$ and $V$ from Q1 are highly coherent with each other. The region of valid values of $q$, $m_H$, and $m_V$ while $\Delta \varphi = 0$ is plotted in Fig.~\ref{fig:cohplot}. It is obvious from Eq.~\eqref{eq:coherence_inequality} that the regions for nonzero $\Delta \varphi$ are subsets of the region in Fig.~\ref{fig:cohplot}.

As is evident from Fig.~\ref{fig:cohplot}, for large enough values of $m_H$ and $m_V$, only a subset of values of $q$ can be attained. The whole range of $q$ from 0 to 1 is accessible only when $0 \leq m_H = m_V \leq 1/\sqrt{2}$. This subset of values is depicted as a dim gray dashed line in Fig.~\ref{fig:cohplot}. In all the other cases, the inequality \eqref{eq:coherence_inequality} restricts the range of input states $\rhoi$ that are consistent with the measured visibilities. For example, when both $m_H$ and $m_V$ are equal to 1, the only allowed value for $q$ is also 1. In such a case, only pure states of the idler photon can give rise to the measured visibilities.

In the general case, when the form of the interaction between the idler photon and the environment is not known, one cannot precisely reconstruct the state of the idler photon. Nevertheless, when we make some physically motivated assumptions on the interactions between the polarization and the environment, we can acquire enough information to reconstruct the mixed state $\rhoi$. In the simplest scenario, when there is no interaction, we can set $\ket{e_H} = \ket{e_V} = \ket{e_\Psi}$ and factor the environment out of the state in Eq.~\eqref{eq:larger_idler_pure}. As a result, the state of the idler photon's polarization is pure, and we recover the results of the previous sections. This scenario is depicted as a red dot in Fig.~\ref{fig:cohplot}.

Another notable situation is when only the vertical polarization interacts with the environment. In contrast, the horizontal polarization is not affected, as is the case in some birefringent materials, and so $\ket{e_H} = \ket{e_\Psi}$ with $m_H = 1$. For such values, the inequality \eqref{eq:coherence_inequality} can be satisfied only when $m_V = q$ and $\Delta \varphi = 0$. This situation is depicted as a red line in Fig.~\ref{fig:cohplot}. Assuming that $T = 1$, the visibility Stokes parameters in Eqs.~\eqref{eq:visc}--\eqref{eq:vis_hv_mixed} reduce to
\begin{eqnarray}
    \visx & = & 2 \, \alpha \, \beta \, q \, \cos(\xi), \\
    \visy & = & 2 \, \alpha \, \beta \, q \, \sin(\xi), \\
    \visz & = & \alpha^2 - (\beta \, q)^2, \\
    \visc & = & \alpha^2 + (\beta \, q)^2.
\end{eqnarray}
The first two formulas are identical to the standard Bloch vector coordinates in Eqs.~\eqref{eq:bloch_coord_x} and \eqref{eq:bloch_coord_y}. The third formula does not match the $z$-coordinate in Eq.~\eqref{eq:bloch_coord_z}, but we can remedy this easily by noting that due to the normalization $\alpha^2 + \beta^2 = 1$, the $z$-coordinate is equal to
\begin{equation}
z = \alpha^2 - \beta^2 = 2 \alpha^2 - 1 = \visz + \visc - 1.
\end{equation}
In this scenario, we can thus reconstruct the polarization state of the idler photon. This special case was considered in Ref.~\cite{fuenzalida2022quantum}, where the constraint $m_H = 1$ was enforced. Obviously, when $m_V = 1$, then analogously $q = m_H$ and a similar discussion can be made.

The last special case we discuss here is when the interaction between the polarization and the environment is symmetric. Specifically, when the state $\ket{e_H}$ deviates from $\ket{e_\Psi}$ in exactly the opposite fashion to the state $\ket{e_V}$. As a result, it holds that (cf. Appendix~\ref{sec:constraint})
\begin{eqnarray}
\ket{e_\Psi} = \frac{1}{\sqrt{2(1+q)}} (\ket{e_H} + \ket{e_V}). \label{eq:symm}
\end{eqnarray}
From there we obtain $m_H = m_V = \sqrt{(1+q)/2}$ and $\Delta \varphi = 0$ and (provided that $T = 1$) the visibility parameters read
\begin{eqnarray}
    \visx & = & \frac{1+q}{2} \, 2 \, \alpha \, \beta \, \cos(\xi), \label{eq:vis_x_symm} \\
    \visy & = & \frac{1+q}{2} \, 2 \, \alpha \, \beta \, \sin(\xi), \label{eq:vis_y_symm} \\
    \visz & = & \frac{1+q}{2} \, (\alpha^2 - \beta^2). \label{eq:vis_z_symm}
\end{eqnarray}
Without any adjustments, we would misinterpret our state as being depolarized by a factor of $(1+q)/2$. In contrast to the most general case of equations \eqref{eq:vis_da_mixed}, \eqref{eq:vis_lr_mixed}, and \eqref{eq:vis_hv_mixed} though, in the present case, we can determine the value of $q$. The zeroth visibility parameter reads
\begin{eqnarray}
    \visc & = & \frac{1+q}{2},
\end{eqnarray}
from where we calculate $q$. At this point, we can renormalize expressions \eqref{eq:vis_x_symm}, \eqref{eq:vis_y_symm}, and \eqref{eq:vis_z_symm} by dividing by $\visc$ and multiplying by $q$ to obtain the standard Bloch coordinates of the mixed state $\rhoi$ in Eq.~  \eqref{eq:mixed_state}. This scenario is depicted as a blue line in Fig.~\ref{fig:cohplot}.

\section{Algebraic properties of visibility Stokes parameters}
\label{sec:algebra}

\begin{figure*}
    \centering
    \includegraphics[width=\linewidth]{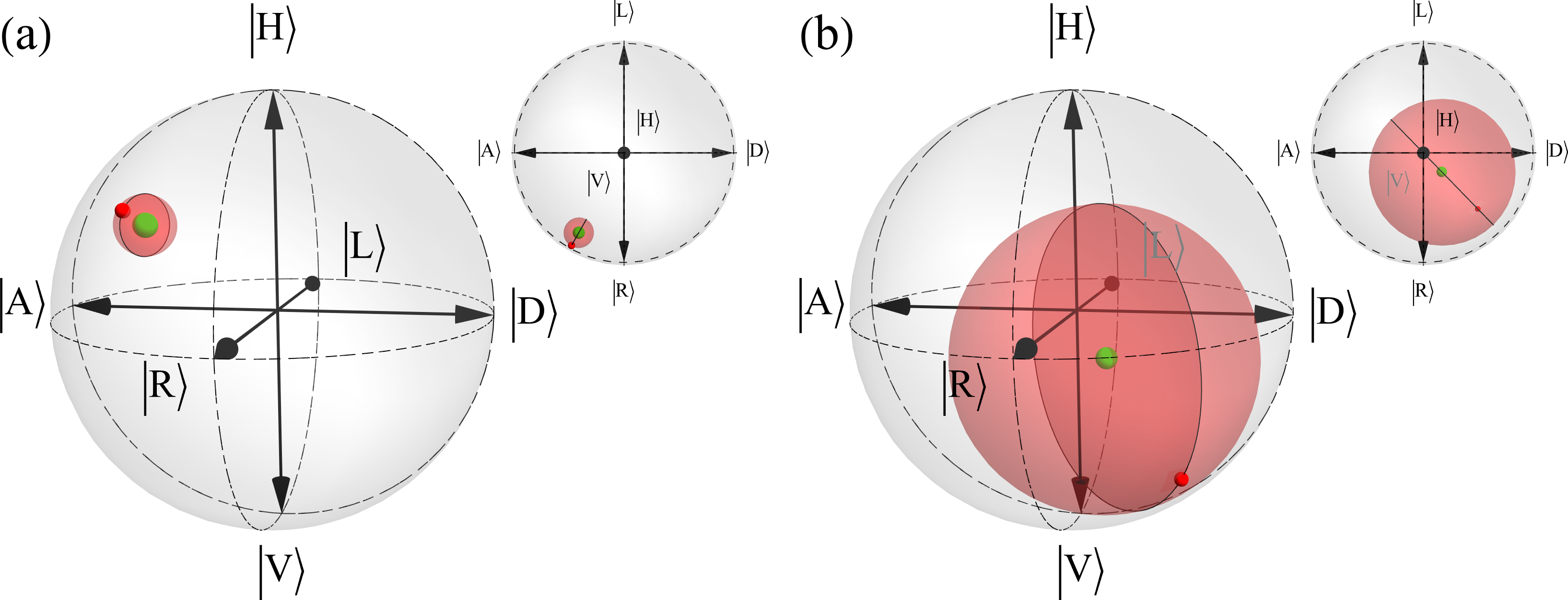}
    \caption{\textbf{From visibilities to polarization states.} One particular visibility Bloch vector $\vec{\stokes{}}$ (green dot) corresponds to many polarization states. All the standard Bloch vectors for quantum states consistent with $\vec{\stokes{}}$ form a ball depicted in red. This ball touches the surface of the Bloch sphere in a single point, highlighted by a red dot. (a) The side view of the Bloch sphere for $\vec{\stokes{}}$ with $|\vec{\stokes{}}| = \visc$ close to one. The top view of the same sphere is shown in the inset. (b) For $\vec{\stokes{}}$ with small $\visc$, the set of relevant states gets larger.}
    \label{fig:vis_to_pol}
\end{figure*}

In analogy to the standard Bloch representation, we can introduce the \emph{visibility Bloch vector} $\vec{\stokes{}}$ as the triple
\begin{eqnarray}
    \vec{\stokes{}} = (\visx, \visy, \visz).
\end{eqnarray}
It can be checked by the direct substitution of their explicit forms in Eqs.~\eqref{eq:vis_da_mixed}--\eqref{eq:vis_hv_mixed} that the three parameters satisfy
\begin{equation}
    \visx^2 + \visy^2 + \visz^2 = \visc^2
    \label{eq:visSq_sc}
\end{equation}
and so the norm of the visibility Bloch vector is equal to the zeroth parameter: $|\vec{\stokes{}}| = \visc$. An important question is how close this visibility Bloch vector is to the actual Bloch vector $\vec{r} = (x, y, z)$. It is straightforward to show from the explicit expressions in Eqs.~\eqref{eq:bloch_coord_x}--\eqref{eq:bloch_coord_z} and Eqs.~\eqref{eq:vis_da_mixed}--\eqref{eq:vis_hv_mixed} together with the inequality in Eq.~\eqref{eq:coherence_inequality}, that
\begin{align}
    (\visx - x)^2 + (\visy - y)^2 + (\visz - z)^2 \leq (1 - \visc)^2
    \label{eq:dist}
\end{align}
and so the Euclidean distance between the two vectors is bounded from above by the \emph{incoherence} $1 - \visc$. All the Bloch vectors consistent with a given $\vec{\stokes{}}$ thus form a ball centered in $\vec{\stokes{}}$ with radius $1 - \visc$, as shown in Fig.~\ref{fig:vis_to_pol}. The ball always touches the surface of the Bloch sphere at a point that corresponds to the pure state of the form $\alpha_E \ket{H} + \sqrt{1 - \alpha_E^2} \exp(i \xi) \ket{V}$, where $\alpha_E = (1/\sqrt{2})\sqrt{1 + \visz / \visc}$ (cf. Appendix~\ref{sec:visc_relations}).

Equation \eqref{eq:visSq_sc} resembles the relation $x^2 + y^2 + z^2 = r^2$ for standard Stokes parameters, where $r \equiv |\vec{r}|$ is the norm of the Bloch vector. This norm relates to the \emph{purity} $\mathscr{P}$ of the corresponding quantum state as $\mathscr{P} = (1 + r^2)/2$. For visibility Stokes parameters, such a relation does not hold. However, from Eqs.~\eqref{eq:visSq_sc} and \eqref{eq:dist} one can easily show that
\begin{align}
    \mathscr{P} \leq (\vec{r} \cdot \vec{\stokes{}}) + 1 - \visc.
    \label{eq:purity_ineq}
\end{align}
One can apply the Cauchy-Schwarz inequality to the inner product in this formula to find the direct relation between the purity $\mathscr{P}$ and the coherence $\visc$. We arrive at the \emph{lower} bound in the form: $\mathscr{P} \geq 1 - 2 \visc(1-\visc)$ as long as $\visc \geq 1/2$ (for $\visc < 1/2$ no special lower bound on $\mathscr{P}$ applies). Analogously, one can derive that $\visc \leq (1 + r) / 2$.

When the inequality \eqref{eq:purity_ineq} is understood as a constraint on varying visibility Stokes parameters while the Bloch vector $\vec{r}$ is fixed, it turns out that this inequality represents a rotational ellipsoid, see Appendix~\ref{sec:visc_relations}. The ellipsoid has its center in point $\vec{r}/2$, its two loci are located in the origin $(0,0,0)$ and point $\vec{r}$, and its semiaxes read $1/2$ and $\sqrt{1-r^2}/2$. Depending on coherence conditions one can thus for a given polarization state of a photon obtain many visibility Bloch vectors and all these vectors form an ellipsoid, see Fig.~\ref{fig:pol_vs_vis}.

Several other algebraic properties of the visibility Stokes parameters can be found in Appendix~\ref{sec:visc_relations}.

\begin{figure*}
    \centering
    \includegraphics[width=\linewidth]{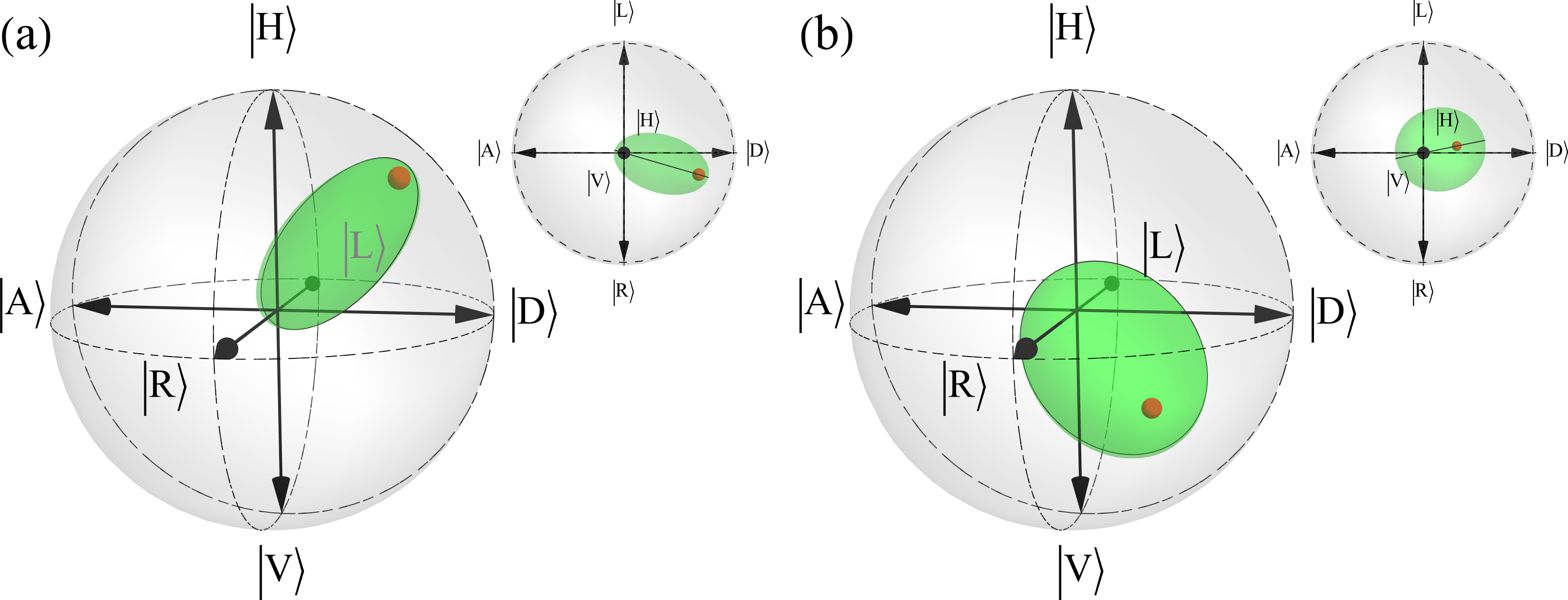}
    \caption{\textbf{From polarization states to visibilities.} One particular quantum state corresponds to many visibility Bloch vectors. The standard Bloch vector $\vec{r}$ of a given quantum state is represented by a red dot, while all the visibility Bloch vectors that are consistent with the state form an ellipsoid depicted in green. (a) The side view of the Bloch sphere for an almost pure polarization state. The top view of the same sphere is shown in the inset. (b) For increasingly more mixed states, the ellipsoid occupies an increasingly larger volume.}
    \label{fig:pol_vs_vis}
\end{figure*}

\section{Measurement operators corresponding to visibility measurements}
\label{sec:quantum_opers}

The standard Stokes parameters can be expressed as expectation values of measurement operators in a given quantum state. In this section, we analogously introduce the operators corresponding to the visibility Stokes parameters. At first, note that from Eq.~\eqref{eq:visk_simple} and the definition of $\ket{\psi'_I}$ it follows that
\begin{eqnarray}
\vis{k}^2 & = & T^2|\braket{\psi'_I}{k^\ast}\ket{e_\Psi}|^2 \nonumber \\
& = & T^2 \braket{\psi'_I}{k^\ast}\ket{e_\Psi} \bra{e_\Psi}\braket{k^\ast}{\psi'_I} \nonumber \\
& = & \bra{\psi_I} \, \visop{k} \, \ket{\psi_{I}}, \label{eq:exp_val}
\end{eqnarray}
where we defined an operator
\begin{equation}
    \visop{k} = T^2 (\mathrm{DM}_I^\dagger \cdot \ket{k^\ast} \bra{k^\ast} \cdot \mathrm{DM}_I) \otimes \ket{e_\Psi} \bra{e_\Psi}.
    \label{eq:vis_op}
\end{equation}
One can therefore express the visibilities as expectation values of specific measurement operators. Following Eqs.~\eqref{eq:s_x}--\eqref{eq:s_z}, these operators can be used to introduce a triple of operators that correspond to the visibility Stokes parameters
\begin{eqnarray}
    \visopx & \equiv & \visop{D} - \visop{A}, \label{eq:oper_x} \\
    \visopy & \equiv & \visop{L} - \visop{R}, \label{eq:oper_y} \\
    \visopz & \equiv & \visop{H} - \visop{V}. \label{eq:oper_z} 
\end{eqnarray}
All of these are Hermitian operators whose form is independent of the idler state $\ket{\psi_I}$. When expressed explicitly, their form turns out to be a tensor product of Pauli matrices and the projector on the state of the second source's environment:
\begin{eqnarray}
    \visopx & = & T^2 (\ket{D_{b'}}\bra{D_{b'}}-\ket{A_{b'}}\bra{A_{b'}})\otimes \ket{e_\Psi}\bra{e_\Psi}, \\
    \visopy & = & T^2 (\ket{L_{b'}}\bra{L_{b'}}-\ket{R_{b'}}\bra{R_{b'}})\otimes \ket{e_\Psi}\bra{e_\Psi}, \\
    \visopz & = & T^2 (\ket{H_{b'}}\bra{H_{b'}}-\ket{V_{b'}}\bra{V_{b'}})\otimes \ket{e_\Psi}\bra{e_\Psi}.
\end{eqnarray}
These operators act on the idler polarization state together with its environment. From Eq.~\eqref{eq:exp_val} it follows that the visibility Stokes parameters (cf. Eqs.~\eqref{eq:s_x}--\eqref{eq:s_z}) coincide with the expectation values of the operators in Eqs.~\eqref{eq:oper_x}--\eqref{eq:oper_z} as
\begin{eqnarray}
    \visx & = & \bra{\psi_I} \ \visopx \  \ket{\psi_I}, \\
    \visy & = & \bra{\psi_I} \ \visopy \  \ket{\psi_I}, \\
    \visz & = & \bra{\psi_I} \ \visopz \  \ket{\psi_I}.
\end{eqnarray}
We have just shown how the visibility measurements that are physically performed on signal photons are expressed as measurement operators acting on the undetected idler photon.\footnote{The visibility operators give information that is in actuality obtained from many copies of the idler photon state, each measured for a different interferometric phase $\phi$.} The fact that this is possible shows how information about the undetected idler photon is accessible without detecting it. The measurement operator associated with the zeroth visibility parameter $\visc$ reads
\begin{equation}
    \visopc = T^2 (\ident \otimes \ket{e_\Psi}\bra{e_\Psi}).
\end{equation}
In the standard tomography, this operator is equal to the identity and does not provide us with any additional information. In our case, though, the expectation value of this operator contains information about the coherence between the idler state and the state of the second source.

Even though the operators $\visopx$, $\visopy$, and $\visopz$ derived above resemble the Pauli operators in Eqs.~\eqref{eq:paulx}--\eqref{eq:paulz}, it is important to emphasize that they do not form a complete measurement in the usual sense. Formally speaking, each visibility Stokes operator $\visopS{k}$ corresponds to two positive measurement operators $\visop{k}$ and $\visop{k^\perp}$. These two operators nevertheless do not satisfy the completeness relation $\sum_k \visop{k} = \ident$. Instead, they sum up to $\visc$, cf. Eq.~\eqref{eq:visc_0}. This can be remedied by expanding each set of visibility operators with an operator $\visop{\mathrm{incoh}} \equiv \ident - \visopc$, quantifying the incoherence between the two sources. Unlike the standard polarization measurement, where e.g. $\hat{\Pi}_H + \hat{\Pi}_V = \ident$ with $\hat{\Pi}_k = \ket{k}\bra{k}$, here we have $\visop{H} + \visop{V} + \visop{\mathrm{incoh}} = \ident$. Let us note that for $T = 1$, the visibility operators, including $\visop{\mathrm{incoh}}$, are projectors.

\section{Post-measurement states}
\label{sec:postmeas_states}

In this section, we briefly discuss the post-measurement state of the idler photon alone. Let us treat for simplicity only pure polarization states $\ket{\psi_{Ib'}}$ \eqref{eq:idler_state}, in which case the pre-measurement state $\ket{\psi}$ \eqref{eq:superpos} can be rewritten into 
\begin{eqnarray}
    \ket{\psi} & \propto & T e^{i \phi} (\mathrm{BS}\ket{\psi_{Sa}})\otimes \ket{\psi'_{I}} \nonumber \\ & + & \sqrt{1-T^2} \, e^{i \phi} (\mathrm{BS}\ket{\psi_{Sa}})\otimes\ket{\psi_{w}} \nonumber \\
    & + & \frac{\rp}{\sqrt{2}} \, (\mathrm{BS}\ket{H_{Sb}})\otimes(\mathrm{DM}_{I}\ket{H_I}) \nonumber \\
    & + & \frac{\rp}{\sqrt{2}} \, (\mathrm{BS}\ket{V_{Sb}})\otimes(\mathrm{DM}_{I}\ket{V_I}).
\end{eqnarray}
Vectors associated with different paths, $a$ and $b$, are orthogonal to each other and so $\braket{\psi_{Sa}}{H_{Sb}} = \braket{\psi_{Sa}}{V_{Sb}} = 0$. Since both the beam splitter and the dichroic mirror are unitary operations, they do not change the orthogonality of vectors, and it is, therefore, not hard to perform a partial trace of the expression above to get the reduced state of the idler photon. We obtain
\begin{eqnarray}
    \rhoi^{\mathrm{post}} & = & \Tr_S (\ket{\psi}\bra{\psi}) \nonumber \\
    & = & \frac{T^2}{1+\rp^2} \ket{\psi'_{I}}\bra{\psi'_{I}} + \frac{\rp^2}{1+\rp^2} \left( \ident / 2 \right) + W,
\end{eqnarray}
where
\begin{eqnarray}
    W & = & \frac{1-T^2}{1+\rp^2}\ket{\psi_w}\bra{\psi_w} \nonumber \\
    & + & \frac{T \sqrt{1-T^2}}{1+\rp^2} \left(\ket{\psi'_I}\bra{\psi_w} + \ket{\psi_w}\bra{\psi'_I} \right)
\end{eqnarray}
is that part of the state that ends up in path $w$. When the transmission of the setup is perfect, and $\rp = 1$, the final state of the idler photon reads
\begin{eqnarray}
    \rhoi^{\mathrm{post}} & = & \frac{1}{2} \ket{\psi'_{I}}\bra{\psi'_{I}} + \frac{1}{2} \left( \ident / 2 \right).
    \label{eq:poststate}
\end{eqnarray}
The final state is a uniform mixture of the maximally mixed state $\ident/2$ and the original pure state $\ket{\psi'_{I}}$ in path $c$ \eqref{eq:idler_state}. Evidently, the post-measurement state of the idler photon still contains some information about its original state of polarization.

\section{Conclusion}
\label{sec:conclusion}

We develop the formalism of visibility Stokes parameters that complements the standard Stokes parameters in the context of coherence-based quantum operations. Specifically, we focus on the technique of quantum state tomography of undetected photons introduced in Ref.~\cite{fuenzalida2022quantum}. The visibility Stokes parameters characterize a quantum system that is not directly measured but whose state information can be extracted via quantum interference using the effect of induced coherence without induced emission \cite{wang1991induced}. This methodology thus profoundly differs from the standard intensity measurements for state reconstruction. The visibility Stokes parameters and the corresponding visibility operators go beyond the problem of quantum state tomography and constitute a bedrock for other quantum information techniques. These operators could enable the translation of single-photon protocols such as the BB84~\cite{bennett_quantum_2014} and the B92~\cite{bennett_quantum_1992} to the realm of optical coherence.

The form of the visibility Stokes parameters is similar to the standard Stokes parameters. Still unlike the latter, which are based on intensity measurements, the former are determined through coherence measurements in the form of visibilities. For the general case of mixed states, there are several relations between the Stokes parameters and their visibility counterparts. In general, a given triple of visibility Stokes parameters can be consistent with many polarization states of the idler photon. Similarly, many triples of visibility Stokes parameters may represent a single polarization state. We discuss possible ways to eliminate this ambiguity and establish a one-to-one relation between the standard and the visibility Stokes parameters.

To analyze the dependence of visibility Stokes parameters on the input state and the physical setup, we thoroughly analyze the environment of the idler photon. The overlap of states of environment for the first and the second source in the setup models the mixedness of the idler photon's state and the mutual coherence between the sources. In this discussion, we assume that the two crystals composing the second source are perfectly coherent and produce a Bell pair of polarization. One could expand this discussion and study the effects of partially coherent crystals. In addition, the concept of mutually unbiased bases is applied to find an efficient implementation of the quantum tomography of undetected photons. This way, not only the form of the resulting formulas is simplified, but one can also monitor counts simultaneously in both outputs in the actual physical setup, not just one as done in Ref.~\cite{fuenzalida2022quantum}.

In conclusion, we introduce a set of quantities akin to the standard Stokes parameters but for quantum systems composed of undetected photons. Whether there are other visibility parameters for other quantum entities remains open. The question of generalizing our results to higher dimensions is left open as well. With this, we expect to pave the way for adapting more state estimation techniques and quantum information protocols for undetected photonic systems.

\section{Acknowledgement}

The authors would like to thank Armin Hochrainer and Mayukh Lahiri for fruitful discussions. J.K. thanks Prof. P. Walther for his hospitality. The financial support by the Austrian Federal Ministry of Labour and Economy, the National Foundation for Research, Technology and Development and the Christian Doppler Research Association is gratefully acknowledged.

\bibliographystyle{unsrt}
\bibliography{ref.bib}

\appendix

\section{Visibilities for pure states}
\label{sec:vispure}

For completeness, in the following we present the general formulas for visibilities in bases $H/V$, $D/A$, and $L/R$, when the relative phase between the two crystals in the second source is equal to $\theta$ (throughout the text we could assume $\theta = 0$ due to the phase calibration), the transmission in the idler photon path is $T$, and the state of the signal photon is of the form
\begin{eqnarray}
    \ket{\psi_{Sa}} = \delta \ket{H_{Sa}} + \varepsilon e^{i \zeta} \ket{V_{Sa}},
\end{eqnarray}
where $\delta^2+\varepsilon^2=1$, $\delta, \varepsilon \ge 0$ and $\zeta \in [0, 2\pi)$. Moreover, let the relative pump power between the first and the second source be equal to $\rp^2$, where $\rp$ is a real number. In such a general case the visibilities are given by
\begin{eqnarray*}
    \vish = \frac{2\sqrt{2} \delta \rp T}{\rp^2+2 \delta^2} \, \alpha, & \quad &
    \visv = \frac{2\sqrt{2} \varepsilon \rp T}{\rp^2+2\varepsilon^2} \, \beta,
\end{eqnarray*}
\begin{equation*}
    \vis{D/A} = \frac{2 \rp T \sqrt{1\pm 2\delta\varepsilon\cos(\zeta)}}{1 + \rp^2 \pm 2 \delta \varepsilon \cos(\zeta)} \,  \sqrt{\frac{1\pm2\alpha\beta\cos(\xi-\theta)}{2}}, \label{eq:visda}
\end{equation*}
\begin{equation*}
    \vis{L/R} = \frac{2 \rp T \sqrt{1 \mp 2\delta\varepsilon\sin(\zeta)}}{1 + \rp^2 \mp 2 \delta \varepsilon \sin(\zeta)} \, \sqrt{\frac{1 \pm 2\alpha\beta\sin(\xi-\theta)}{2}}. \label{eq:visrl}
\end{equation*}
We recover the formulas \eqref{eq:vis_hv}--\eqref{eq:vis_rl} from these general expressions for $T = 1$, $\rp = 1$, $\theta = 0$ and $\delta = \varepsilon = 1/\sqrt{2}$.

\section{Visibilities for mixed states}
\label{sec:vismixed}

The formula for the detection probability in Eq.~\eqref{eq:interference} can be recast into
\begin{multline}
    \bra{\psi} \hat{\Pi}_j \, \ket{\psi} = 2 \, \rp \, T \mathcal{N}^2 \, \mathrm{Re}(z_j) \cos(\phi) \\ + 2 \, \rp \, T \mathcal{N}^2 \, \mathrm{Im}(z_j) \sin(\phi) + \mathcal{N}^2 c_j,
    \label{eq:interference_2}
\end{multline}
which is of the form $f(\phi) = A \cos(\phi) + B \sin{(\phi)} + C$ for real constants $A$, $B$, and $C$. One can always find a non-negative $D$ and real $\omega$ such that $A = D \sin(\omega)$ and $B = D \cos(\omega)$. The original expression can then be rewritten into $f(\phi) = C + D \sin(\phi + \omega)$, from which it is evident that the visibility is equal to $D/C$. Since $A^2 + B^2 = D^2$ we get that the visibility equals $\sqrt{A^2 + B^2}/C$. When we plug the explicit forms of $A$, $B$, and $C$ into this formula, we obtain the expression in Eq.~\eqref{eq:vis_j}.

\section{Visibilities for unbiased signal states}
\label{sec:vis_mub}

In this section we derive the form of visibilities in Eq.~\eqref{eq:visk_simple} from the general formula in Eq.~\eqref{eq:vis_j}. First, note that the state produced by the second source, Eq.~\eqref{eq:source_2_psi}, is the tensor product of the Bell state $\ket{\Phi^+}$ and the environmental state $\ket{e_\Psi}$. The Bell state exhibits a unitary invariance of the form $U \otimes U^\ast \ket{\Phi^+} = \ket{\Phi^+}$ for an arbitrary one-qubit unitary $U$ and its complex conjugate $U^\ast$. If one projects the first qubit on state $\ket{k}$, the post-measurement state of the second qubit reads
\begin{align}
    \braket{k}{\Phi^+} & = \bra{H} (U^\dagger \otimes \ident) \ket{\Phi^+} \nonumber \\
    & = \bra{H} (\ident \otimes U^\ast)(U^\dagger \otimes U^T) \ket{\Phi^+} \nonumber \\
    & = \bra{H} (\ident \otimes U^\ast) \ket{\Phi^+} \nonumber \\
    & = (1/\sqrt{2}) U^\ast \ket{H}, \label{eq:post_meas_inv}
\end{align}
where $\ket{k} = U \ket{H}$ for a specific unitary $U$, and where we used the unitary invariance for $U^\dagger$.

We model the symmetric beam splitter by a Hadamard matrix. If we fix the projector $\hat{\Pi}_j$ to act (initially) on path $e$, it is easy to show that the brakets in the definition of $c_j$ in Eq.~\eqref{eq:c_j} reduce to $\bra{\psi'_{S}} \hat{\Pi}_j \ket{\psi'_{S}} = \bra{\psi_{S}} \mathrm{BS}^\dagger \hat{\Pi}_j \mathrm{BS} \ket{\psi_{S}} = (1/2) \bra{\psi_{S}} \hat{\Pi}_j \ket{\psi_{S}}$ and
$\bra{\Psi'} \hat{\Pi}_j \ket{\Psi'} = \bra{\Phi^+} \mathrm{BS}^\dagger \hat{\Pi}_j \mathrm{BS} \ket{\Phi^+} = (1/2) \bra{\Phi^+} \hat{\Pi}_j \ket{\Phi^+}$. Similarly, $z_j$ in Eq.~\eqref{eq:z_j} reduces to $\bra{\psi'_{S}}\otimes\bra{\psi'_{I}} \hat{\Pi}_j \ket{\Psi'} = \bra{\psi_{S}} \mathrm{BS}^\dagger \hat{\Pi}_j \mathrm{BS} \otimes \braket{\psi_{I}}{\Psi} = (-1/2) \bra{\psi_{S}} \hat{\Pi}_j \otimes \braket{\psi_{I}}{\Psi}$.

From Eq.~\eqref{eq:post_meas_inv} we further get $\bra{\Phi^+} \hat{\Pi}_j \ket{\Phi^+} = \braket{\Phi^+}{j} \braket{j}{\Phi^+} = 1/2$. As mentioned in the main text, we set the signal's state $\ket{\psi_S}$ to be mutually unbiased with the projection vector $\ket{j}$, where $\hat{\Pi}_j = \ket{j}\bra{j}$, that is $\braket{\psi_S}{j} = (1/\sqrt{2}) \exp(i \varphi_j)$ for some real $\varphi_j$. From there it follows that $ \bra{\psi_{S}} \hat{\Pi}_j \ket{\psi_{S}} = 1/2$ and $\bra{\psi_{S}} \hat{\Pi}_j \otimes \braket{\psi_{I}}{\Psi} = (1/\sqrt{2}) \exp(i \varphi_j) \bra{j} \braket{\psi_{I}}{\Phi^+}\ket{e_\Psi} = (1/2) \exp(i \varphi_j) \bra{\psi_{I}} U_j^\ast \ket{H} \ket{e_\Psi}$, where $\ket{j} = U_j \ket{H}$ and where we used Eq.~\eqref{eq:post_meas_inv}.

Taken all these simplifications into account, we can recast the form of $c_j$ and $|z_j|$ into $c_j = (1/4)(1 + \rp^2)$ and $|z_j| = (1/4) |\braket{\psi_{I}}{j^\ast}\ket{e_\Psi}|$, where we define $\ket{j^\ast} = U_j^\ast \ket{H}$. The ratio of these two quantities determines the visibility, for which we obtain
\begin{equation}
    \vis{j} = 2 \, \rp \, T \frac{|\braket{\psi_{I}}{j^\ast}\ket{e_\Psi}|}{1 + \rp^2}.
\end{equation}
As is easy to see, when setting $\rp = 1$, this formula reduces to Eq.~\eqref{eq:visk_simple} as we wanted to show.

\section{Notable properties of visibility Stokes parameters}
\label{sec:visc_relations}

The quantum states are normalized such that the standard zeroth Stokes parameter is always unity for both pure and mixed states. If we follow the same logic and divide all the parameters in Eqs.~\eqref{eq:vis_da_mixed}--\eqref{eq:vis_hv_mixed} as well as $\visc$ in Eq.~\eqref{eq:visc} by $\visc$, provided that $\visc \neq 0$, these turn into
\begin{eqnarray}
    \overline{\visx} \equiv \visx/\visc & = & 2 \, \overline{\alpha} \, \overline{\beta} \, \cos(\overline{\xi}), \label{eq:sxdiv} \\
    \overline{\visy} \equiv \visy/\visc & = & 2 \, \overline{\alpha} \, \overline{\beta} \, \sin(\overline{\xi}), \label{eq:sydiv} \\
    \overline{\visz} \equiv \visz/\visc & = & \overline{\alpha}^2 - \overline{\beta}^2, \label{eq:szdiv} \\
    \overline{\visc} \equiv \visc/\visc & = & 1, \label{eq:scdiv}
\end{eqnarray}
where $(\overline{\alpha},\overline{\beta})$ is a unit vector with
\begin{eqnarray}
    \overline{\alpha} & = & \frac{\alpha \, m_H}{\sqrt{\alpha^2 \, m^2_H + \beta^2 \, m^2_V}}, \quad \overline{\beta} = \frac{\beta \, m_V}{\sqrt{\alpha^2 \, m^2_H + \beta^2 \, m^2_V}}, \nonumber \\
    \overline{\xi} & = & \xi + \Delta \varphi.
\end{eqnarray}
The normalized visibility Stokes parameters in Eqs.~\eqref{eq:sxdiv}--\eqref{eq:scdiv} thus formally correspond to Stokes parameters of a pure state $\ket{\overline{\psi}} = \overline{\alpha} \ket{H} + \overline{\beta} \exp(i \overline{\xi}) \ket{V}$. The transmission coefficient $T$ is no longer present in the normalized formulas, but the problems mentioned in the main text remain. From this discussion it is obvious that the visibility Stokes parameters do not directly hold any information about the purity of the measured state. For a given triple of (non-normalized) visibility Stokes parameters, there is exactly one pure state $\ket{\overline{\psi}}$ determined by the corresponding normalized parameters. From the normalization condition $\overline{\alpha}^2 + \overline{\beta}^2 = 1$ and Eq.~\eqref{eq:szdiv} it follows directly that for this state $\overline{\alpha} = (1/\sqrt{2})\sqrt{1 + \visz / \visc}$.

It is easy to see that Eq.~\eqref{eq:purity_ineq} is invariant under the simultaneous rotation of both $\vec{r}$ and $\vec{\stokes{}}$. We can thus rotate the coordinate system such that $\vec{r} = (r,0,0)$, upon which rotation the visibility Bloch vector turns into $(\visx',\visy',\visz')$. The inequality \eqref{eq:purity_ineq} then turns into
\begin{equation}
    \visc \leq r \visx' + (1-r^2)/2
\end{equation}
as the norm $\visc$ of the visibility Stokes vector remains the same. Since $\visc \geq 0$, both sides of this inequality are non-negative and the inequality is preserved when one takes the square of both sides. When one does it and simplifies the result, one can recast the final expression into the form
\begin{equation}
    \frac{(\visx' - r/2)^2}{(1/2)^2} + \frac{(\visy')^2}{(\sqrt{1-r^2}/2)^2} + \frac{(\visz')^2}{(\sqrt{1-r^2}/2)^2} \leq 1,
\end{equation}
which describes a rotational ellipsoid with its center in point $\vec{r}/2$ and with semiaxes $1/2$ and $\sqrt{1-r^2}/2$, while the two loci are situated in the origin $(0,0,0)$ and point $\vec{r}$. These properties are preserved when one expresses the ellipsoid in the original variables $\vec{\stokes{}}$.

Let us mention some neat algebraic relations for $\visc$. Namely,
\begin{align}
    \visd^2 + \visa^2 + \visl^2 + \visr^2 + \vish^2 + \visv^2 & = 3 \visc, \label{eq:visc_f1} \\
    \visd^4 + \visa^4 + \visl^4 + \visr^4 + \vish^4 + \visv^4 & = 2 \visc^2, \label{eq:visc_f2} \\
    \visd^2 \visa^2 + \visl^2 \visr^2 + \vish^2 \visv^2 & = (1/2) \visc^2. \label{eq:visc_f3}
\end{align}
As the left-hand sides contain visibilities measured in all three standard bases, the value of $\visc$ thus obtained might be less sensitive to experimental imperfections. Formula \eqref{eq:visc_f1} follows directly from the threefold use of the definition of $\visc$ in Eq.~\eqref{eq:visc_0}. When we square both sides of Eq.~\eqref{eq:visc_0} and analogously to Eq.~\eqref{eq:visc_f1} sum up its three versions, we arrive at
\begin{equation}
    (\vis{D}^2 + \vis{A}^2)^2 + (\vis{L}^2 + \vis{R}^2)^2 + (\vis{H}^2 + \vis{V}^2)^2 = 3 \visc^2.
\end{equation}
Furthermore, from Eq.~\eqref{eq:visSq_sc} we get
\begin{equation}
    (\vis{D}^2 - \vis{A}^2)^2 + (\vis{L}^2 - \vis{R}^2)^2 + (\vis{H}^2 - \vis{V}^2)^2 = \visc^2.
\end{equation}
When we sum up these last two equations, we immediately obtain Eq.~\eqref{eq:visc_f2}. When we instead subtract the second equation from the first, we get Eq.~\eqref{eq:visc_f3}.

In Eqs.~\eqref{eq:exp_val} and \eqref{eq:vis_op} the visibility operator is introduced. It corresponds to the visibility of the signal photon rate detected in polarization given by a unit vector $\ket{k}$. This vector forms an orthonormal basis with vector $\ket{k^\perp}$ and so $\ket{k}\bra{k} + \ket{k^\perp}\bra{k^\perp} = \ident$. Using this completeness relation one can follow the derivation in Eq.~\eqref{eq:exp_val} to arrive at
\begin{equation}
    \vis{k^\perp}^2 = \bra{\psi_I} \, (T^2 (\ident \otimes \ket{e_\Psi} \bra{e_\Psi}) - \visop{k}) \, \ket{\psi_{I}}.
\end{equation}
This formula evaluates to $\vis{k^\perp}^2 = T^2(\alpha^2 m_H^2 + \beta^2 m_V^2) - \vis{k}^2$, from where one immediately sees that
\begin{equation}
    \vis{k}^2 + \vis{k^\perp}^2 = T^2(\alpha^2 m_H^2 + \beta^2 m_V^2) \equiv \visc
\end{equation}
for any orthonormal basis $\{ \ket{k}, \ket{k^\perp} \}$. Similarly, it is straightforward to show that
\begin{equation}
    \vis{k}^2 - \vis{k^\perp}^2 = 2 \vis{k}^2 - \visc
\end{equation}
for any such basis. For the three Pauli bases this expression thus gives us the corresponding visibility Stokes parameters.

\section{Constraints on coherence parameters}
\label{sec:constraint}

The three coherence terms $q$, $m_V$ and $m_V$ are defined as scalar products of three environmental states $\ket{e_H}$, $\ket{e_V}$, and $\ket{e_\Psi}$ introduced in Eqs.~\eqref{eq:overlap} and \eqref{eq:mHmV}. Without loss of generality we can write
\begin{equation}
    \ket{e_\Psi} = \gamma_H \ket{e_H} + \gamma_V \ket{e_V} + \gamma_\perp \ket{e_\perp}
\end{equation}
for some complex numbers $\gamma_H$, $\gamma_V$, and $\gamma_\perp$ and a vector $\ket{e_\perp}$, for which $\braket{e_H}{e_\perp} = \braket{e_V}{e_\perp} = 0$. The norm of $\ket{e_\Psi}$ reads
\begin{equation}
    \braket{e_\Psi}{e_\Psi} = |\gamma_H|^2 + |\gamma_V|^2 + |\gamma_\perp|^2 + 2 q \, \mathrm{Re}(\gamma_H^\ast \gamma_V)
\end{equation}
and has to be equal to unity. From this condition we get
\begin{equation}
    |\gamma_H|^2 + |\gamma_V|^2 + 2 q \, \mathrm{Re}(\gamma_H^\ast \gamma_V) = 1 - |\gamma_\perp|^2 \leq 1.
    \label{eq:norm_epsi}
\end{equation}
It further holds that
\begin{eqnarray}
        m_H e^{i \varphi_H} \equiv \braket{e_H}{e_\Psi} & = & \gamma_H + \gamma_V \, q, \\
        m_V e^{i \varphi_V} \equiv \braket{e_V}{e_\Psi} & = & \gamma_H \, q + \gamma_V,
\end{eqnarray}
from where one gets
\begin{eqnarray}
    \gamma_H & = & (m_H e^{i \varphi_H} - q \, m_V e^{i \varphi_V})/(1 - q^2), \\
    \gamma_V & = & (m_V e^{i \varphi_V} - q \, m_H e^{i \varphi_H})/(1 - q^2).
\end{eqnarray}
By plugging these expressions into the inequality \eqref{eq:norm_epsi} and simplifying the result one arrives at
\begin{equation}
    1-\left(q^2+m_H^2+m_V^2\right) + 2 \, q \, m_H \, m_V \cos(\Delta \varphi) \geq 0
    \label{eq:coherence_inequality_app}
\end{equation}
with $\Delta \varphi = \varphi_H - \varphi_V$, as we wanted to show. From there we further obtain the inequality
\begin{equation}
    1-\left(q^2+m_H^2+m_V^2\right) + 2 \, q \, m_H \, m_V \geq 0,
    \label{eq:coherence_inequality_coarse}
\end{equation}
which is independent of $\Delta \varphi$. This same inequality can also be derived from general considerations for density matrices, where one requires the matrix to be positive semi-definite. Similar discussion was done in the supplementary of Ref.~\cite{fuenzalida2022quantum}. When we identify $q$ and $m_V$ with the coherence parameters $\mathscr{I}$ and $\mathscr{L}$ from Ref.~\cite{fuenzalida2022quantum} and set $\mathscr{L} = \mathscr{L}'$ and $m_H = 1$, then inequality \eqref{eq:coherence_inequality_app} can be satisfied only when $\mathscr{I} = \mathscr{L}$. Exactly this condition was also obtained in Ref.~\cite{fuenzalida2022quantum}.

Note that the inequalities above contain only inner products of environmental states and are thus invariant under the unitary evolution acting on the environment. This way, we remove the arbitrariness in our original choice of $\ket{e_H}$, $\ket{e_V}$, and $\ket{e_\Psi}$.

When the environment is known to be only two-dimensional, e.g. when the environment is artificially set by an experimenter, the inequality \eqref{eq:coherence_inequality_coarse} turns into a quadratic equation, which we can solve for $q$ obtaining $q_{\pm} = m_H m_V \pm \sqrt{1 + m_H^2 m_V^2 - m_H^2 - m_V^2}$. For $m_H = 1$ (or $m_V = 1$), the expression under the square root vanishes and we are left with a unique solution $q_+ = q_- = q = m_V$ (or $q = m_H$) that allows for a perfect reconstruction of the mixed state $\rhoi$. This condition was enforced in Ref.~\cite{fuenzalida2022quantum} and the reconstruction of pure idler state is obviously a special case of this condition with $m_H = m_V = 1$. Let us recall that $q$ has to lie in the interval $[0,1]$. The ``plus'' solution $q_+$ complies with this requirement, while the ``minus'' solution $q_-$ complies as long as $m_H^2 + m_V^2 \geq 1$. For the region with $m_H^2 + m_V^2 < 1$ we thus obtain a unique solution $q_+$. If it further holds that $m_H = m_V$, the ``plus'' solution reduces to $q_+ = 1$, while the ``minus'' solution reads $q_- = 2 m_H^2 - 1$ as long as $1/\sqrt{2} \leq m_H = m_V \leq 1$. This last condition is exemplified by the state in Eq.~\eqref{eq:symm} in the main text.

\end{document}